\documentclass[preprint,amsmath,amssymb,aps,prd,showpacs]{revtex4-1}

\usepackage{epsfig}
\usepackage{dcolumn}
\usepackage{bm}
\usepackage{tikz}
\usepackage{graphicx, graphics, color}
\usepackage{dcolumn}
\usepackage{bm}
\usepackage{hyperref}
\usepackage[mathlines]{lineno}
\usepackage{float}
\usepackage{subfig}

\newcommand{\be}{\begin{equation}}
\newcommand{\ee}{\end{equation}}
\newcommand{\ben}{\begin{eqnarray}}
\newcommand{\een}{\end{eqnarray}}

\newcommand{\cZ}{{\cal Z}}
\newcommand{\cD}{{\cal D}}

\newcommand{\cO}{{\cal O}}

\newcommand{\cL}{{\cal L}}

\newcommand{\p}{\partial}

\newcommand{\tA}{\tilde A}
\newcommand{\tF}{\tilde F}

\newcommand{\tZ}{\tilde Z}

\newcommand{\ep}{\epsilon}

\newcommand{\ga}{\gamma}

\newcommand{\Dsl}{{\slash \negthinspace \negthinspace \negthinspace \negthinspace  D}}
\newcommand{\psl}{{\slash \negthinspace \negthinspace \negthinspace \negthinspace  \p}}
\newcommand{\tAsl}{{\slash \negthinspace \negthinspace \negthinspace \negthinspace  {\tilde A}}}
\newcommand{\tBsl}{{\slash \negthinspace \negthinspace \negthinspace \negthinspace  {\tilde B}}}

\newcommand{\tB}{{\tilde B}}

\newcommand{\te}{\tilde e}

\graphicspath{{figs/}}
\graphicspath{{../figs/}}

\newcommand{\avr}[1]{{\left\langle #1 \right\rangle}}

\begin{document} 

\title{Conformal anomaly transport induced by dark photon}

\author{Marek Rogatko} 
\email{rogat@kft.umcs.lublin.pl}
\author{Karol I. Wysoki\'nski}
\email{karol.wysokinski@mail.umcs.pl}
\affiliation{Institute of Physics, 
Maria Curie-Sk\l{}odowska University, 
pl.~Marii Curie-Sk\l{}odowskiej 1,  20-031 Lublin,  Poland}


\date{\today}

\begin{abstract}
We have considered the problem of the influence of inhomogeneity of gravitational field on transport effects predicted by the field theory describing
massless Dirac fermions in the Maxwell and {\it dark matter} background.
As a model of {\it dark sector} one takes into account {\it dark photon} model,  where the {\it hidden sector} is described by the auxiliary $U(1)$-gauge field 
coupled to the {\it visible sector}.  Elaborating the model we restrict our considerations to the case when Weyl type conformal transformation slightly differs from the Minkowski spacetime.
This assumption simplifies the calculations and enables us not to use complicated methods of the quantum field theory in the curved background.
The resulting currents stemming both from {\it visible} and {\it dark sectors} are proportional to the adequate beta functions appearing in the elaborated systems.
For charge-less {\it dark sector} we predict corrections to the scale conductivities in both sectors: linear in $\alpha$ in the dark sector and quadratic in the visible one.

\end{abstract}

\maketitle
\flushbottom

\section{Introduction}
\label{sec:intro}
Symmetries play an important role in modern physics. On the classical level, with each symmetry of the studied system,
one can connect the corresponding conservation law, as dictated by the celebrated Noether theorem~\cite{neu11}.


However, it turns out that in some situations the quantum effects make the symmetry of the classic Lagrangian incompatible with quantum theory, 
then one speaks about anomalies~\cite{ber96}. The anomalies have first been discussed in quantum field theory~\cite{shi91} and recently became an 
active field of research in condensed matter physics~\cite{lan16}. This is related to the discoveries of materials with Dirac like energy spectrum in two dimensional graphene sheets 
and the two-dimensional "metalic-xenes", and latter in three dimensional Dirac and/or Weyl semimetals. 
Here, the quantum anomalies are visible by their effect on quantum transport~\cite{lan16}.

The problem of anomalous transport phenomena acquires recently great attention. 
 Quantum effects break classically preserved symmetries and lead to the non-conservations of certain currents. For instance the breaking of axial symmetry of chiral fermions causes emergence of axial current at quantum level \cite{shi91}, or anomalous transport like chiral separation effect \cite{son04}-\cite{met05}, chiral magnetic effect \cite{fuk08}-\cite{vil80} etc.
It also happens that axial anomaly is 
connected with chiral vortical effect \cite{vil79}-\cite{son09}.

In a curved background one has to do with axial-gravitational anomaly \cite{lan11}. On the other hand, the anomaly connected with scale transformation of flat metric (the so-called Weyl scale transformation) has attracted much attention in hydrodynamics, see, e.g., \cite{hyd}. 
The scale anomaly leading to the new transport effects emerging in an external electromagnetic field, on an inhomogeneous gravitational background
has been discussed in Ref. \cite{che16}. It was shown that in an inflating background, the QED scale anomaly~\cite{che16} is responsible for 
generation of additional effects, like electric current
being opposite to the background one. The anomalous currents in the elaborated theories are  proportional to the adequate $\beta$-functions.

The anomalous currents can in principle be measured in condensed matter systems. For example, in Ref. \cite{che18} it has been revealed that in Dirac/Weyl semimetals a conformal Weyl anomaly leads to a bulk current perpendicular to a temperature gradient and direction of magnetic background field. 
Dirac and Weyl semimetals having relativistic energy spectrum  turn out to be a fruitful and important platform to observe in laboratory various effects, earlier predicted or expected in high energy physics. 
These include chiral and other anomalies. The recent discussion of this subject can be found in the review~\cite{che22}.

The purpose of this paper is to analyze the effect of quantum scale anomaly in the model of {\it dark photon}, representing the {\it dark sector}.
Up to now we possess no direct evidence of {\it dark sector} in Earth experiments. The Standard Model also fails to explain the existence of the appropriate candidate for the additional {\it hidden sector}.
The possible solution of this problem is to look for the new physics beyond
the Standard Model. One of a very reasonable candidates for extension of the Standard Model is the idea of {\it dark photon}. It is described by $U(1)$-gauge field
which is coupled to the ordinary Maxwell one by the so-called {\it kinetic mixing term}  \cite{hol86,cap21}. The model in question is also justified by
the contemporary unification scheme  \cite{ach16}, where we have the mixing portals coupling Maxwell 
and auxiliary gauge fields charged under their own groups. From theoretical point of view  {\it dark photons} might be produced 
during inflationary fluctuations, reheating, exotic particle decays, axion oscillations and even from topological defects, like cosmic strings \cite{gra16}-\cite{cstrings}.

On the other hand,  several astrophysical phenomena can be explained by the concept of {\it dark photon}. For instance
$511~ keV$ gamma rays \cite{jea03}, excess of the positron cosmic ray flux in galaxies \cite{cha08}, an anomalous monochromatic $3.56~ keV$ X-ray line in the 
spectrum of some galaxy clusters \cite{bub14}. Astrophysical and laboratory experiments \cite{fil20}, 
are dedicated to establish the range of values for {\it dark photon} - Maxwell field coupling constant and the mass of the hidden photon. They are connected with the
studies of gamma rays emissions from dwarf galaxies \cite{ger15},
inspection of dilaton-like coupling to photons caused by ultra-light {\it dark matter} \cite{bod15} 
 fine structure constant oscillations \cite{til15}, {\it dark photon} emission during supernova event \cite{cha17}, 
 electron excitation measurements in CCD-like detector \cite{sensei}, the search for a {\it dark photon} in $e^+ e^-$ collisions at BABAR experiment \cite{lee14},
 measurements of the muon anomalous effect \cite{dav11}.
 
In our paper
we have revealed that the scale anomaly, studied below, leads to the anomalous contributions to the current induced by external electro-magnetic $U(1)$-gauge fields
in {\it visible} and {\it hidden} sectors. In what follows we suppose that the system in question will be at zero temperature.
This kind of phenomena were studied for the first time in case of massless Dirac $U(1)$-gauge system in Ref. \cite{che16}, and the present paper can be viewed as 
its direct generalisation to the case of two interacting gauge fields.

The organization of the paper is as follows. In the next Sec. II we describe the basic features of  {\it dark photon} model, while in Sec. III the massless Dirac field influenced by the
{\it hidden sector} has been elaborated. Sec. IV is devoted to a scale factor induced current. We have found that the scale current is affected by $\beta$-functions of the effective charges
appearing in the theory. In the case when the spatial derivative of a scale factor is equal to zero, one obtains the so-called scale electric dark matter effect. On the other hand,
when the time derivative of a conformal factor vanishes we get scale magnetic dark matter effect. In Sec. V we present two examples of a scale factor current. One induced by weak gravitational wave passing through the system, the other describing the influence of homogeneous expansion of the Universe on the current.
In Sec. VI we concluded our investigations. In appendix we sketch the derivation of $\beta$-functions for {\it dark photon} theory.
In the paper we use spacetime metric tensor of the following signature, 
$\eta_{\mu \nu} = diag( 1,-1,-1,-1)$.


\section{The Model of dark photon} \label{sec:darkphoton}
In this section we shortly describe the {\it dark photon} model and propose a transformation of underlying gauge fields in order to simplify 
the underlying action, i.e., dispose of {\it kinetic mixing} term. In the process of this we obtain new gauge fields and effective charges, being the mixture of the starting ones, with adequate factors
comprising $\alpha$-coupling constant. Next, one derives the equation of motion for charged particle affected by two gauge fields.

The action describing two coupled, massless gauge field is given by
\be
S_{M-dark~ photon} = \int  d^4x  \Big(
- \frac{1}{4}F_{\mu \nu} F^{\mu \nu} - \frac{1}{4}B_{\mu \nu} B^{\mu \nu} - \frac{\alpha}{4}F_{\mu \nu} B^{\mu \nu}
\Big),
\label{ac dm}
\ee  
where $\alpha$ is
the kinetic mixing parameter connected with the interaction strength between {\it dark photon} and Maxwell one.
As far as the value of the constant in question is concerned, recently
the new limit for it, $\alpha =1.6 \times
10^{-9}$ and the mass range of {\it dark photon}
$2.1\times 10^{-7} - 5.7 \times10^{-6} eV$, have been found in Ref. \cite{fil23}. On the other hand,
quantum limited amplification enable for the first time, to establish the kinetic mixing coupling constant to $10^{-12}$ level for majority of {\it dark photon} masses \cite{ram23}. 
An upper bound on 
kinetic mixing parameter
 $\alpha < 0.3-2 \times 10^{-10}$ (at 95 percent confidence level) has been found in \cite{kot23}.

In the massless kind of {\it dark photon} theory,
for the case of $U(1)$-gauge groups, one can propose the transformation 
\ben \label{transA}
\tA_\mu &=& \frac{\sqrt{2 -\alpha}}{2} \Big( A_\mu - B_\mu \Big),\\ \label{transB}
\tB_\mu &=& \frac{\sqrt{2 + \alpha}}{2} \Big( A_\mu + B_\mu \Big).
\een
 which enables to get rid of the kinetic mixing term from the action (\ref{ac dm}), $\alpha$-kinetic mixing parameter survives in the modified definitions of gauge fields. 
  Importantly, it also enters the coupling constants between {\it dark sector} and matter. This has important consequences and will discussed later.
 As a result one leaves with only modified gauge fields, i.e., 
\be
 F_{\mu \nu} F^{\mu \nu} +
B_{\mu \nu} B^{\mu \nu} +  \alpha F_{\mu \nu} B^{\mu \nu}
\Longrightarrow
 \tF_{\mu \nu} \tF^{\mu \nu} +
\tB_{\mu \nu} \tB^{\mu \nu},
\label{duality}
\ee
where we set $\tF_{\mu \nu} = \p_\mu \tA_{\nu}  - \p_\nu \tA_\mu$, 
and respectively $\tB_{\mu \nu} = \p_\mu \tB_{\nu}  - \p_\nu \tB_\mu$.
Just the action can be rewritten as
\be
S_{m} = \int d^4x  \Big(
- \frac{1}{4}\tF_{\mu \nu} \tF^{\mu \nu} - \frac{1}{4}\tB_{\mu \nu} \tB^{\mu \nu}
\Big).
\label{vdc}
\ee
 Variation of the action (\ref{vdc}) with respect to $\tA_\mu$ and $\tB_\mu$ reveals the following equations of motion for Maxwell {\it dark matter} system, without currents:
\be
\p_{\mu} \tF^{\mu \nu } = 0, \qquad \p_{\mu} \tB^{\mu \nu } = 0.
\label{fb}
\ee

 Having in mind relations (\ref{vdc}) and (\ref{fb}), we shall search for the equation of motion of charged particle in the background of the modified gauge fields. 
 However, the consistency of the approach requires the appropriate redefinitions of the charges coupled to original fields,
where  for transformed charges are defined as
\ben \label{cA}
\te_A &=& \frac{\sqrt{2 -\alpha}}{2} \Big( e - e_d \Big),\\ \label{cB}
\te_B &=& \frac{\sqrt{2 + \alpha}}{2} \Big( e + e_d \Big).
\een
In the above equations $e$ stands for the Maxwell charge, while $e_d$ is connected with {\it dark sector} one. 

The usual procedure in {\it dark photon} massless model \cite{cap21}, \cite{fab21}-\cite{an13} is to  treat $\alpha$ as a constant parameter.
In {\it dark photon} model it is generally assumed that the kinetic mixing parameter corresponds to the case of zero momentum transfer and remains constant with the increase of the momentum transfer (we have no momentum transfer connected with $\alpha$-kinetic mixing term).

Although the {\it kinetic mixing} term can be removed by a field definition in the massless case, relations (\ref{transA})-(\ref{transB}), the parameter $\alpha$ generally remains physical because
it enters the effective coupling constant.
In our work we adopt the simplifying assumption that running of $\alpha$ is neglected in the considered regime.

\section{Massless Dirac fermions in the field of dark photon}

 Taking into account the field transformations
we shall consider {\it dark sector} coupled to massless Dirac fermion field $\psi$ given by the action
\be
S = \int d^4x  \Big(
- \frac{1}{4} \tF_{\mu \nu} \tF^{\mu \nu} -  \frac{1}{4}\tB_{\mu \nu} \tB^{\mu \nu} + i~{\bar \psi}  ~\Dsl \psi
\Big),
\label{S-action}
\ee
where $\Dsl = \ga^\mu D_\mu$, and the covariant derivative is written as follows:
\be
D_\mu = \p_\mu + i ~( \te_A \tA_\mu + \te_B \tB_\mu).
\ee
The form of the covariant derivative envisages the fact that both {\it visible} and {\it hidden} sectors act on the massless fermion field.

The energy momentum tensor $T_{\mu \nu}$
for the theory in question yields
\be
T_{\mu \nu} = T_{\mu \nu}(\tA_\mu) + T_{\mu \nu}(\tB_\mu) + T_{\mu \nu}(\psi),
\ee
where the adequate components, connected with the gauge and fermion fields are given by
\ben
 T_{\mu \nu}(\tA_\mu) &=& -  \tF_{\mu \ga} \tF_\nu {}^{\ga} + \frac{1}{4} \eta_{\mu \nu} \tF^2,\\
 T_{\mu \nu}(\tB_\mu) &=& -  \tB_{\mu \ga} \tB_\nu {}^{\ga} + \frac{1}{4} \eta_{\mu \nu} \tB^2,\\
  T_{\mu \nu}(\psi) &=& i~{\bar \psi} \Big( \ga_\mu D_\nu \psi +  \ga_\nu D_\mu \psi \Big) - \eta_{\mu \nu} i~{\bar \psi}~ \Dsl \psi.
  \een
	
It is worth mentioning that at a classical level the elaborated theory is invariant under redefinition of the absolute length/energy scale.
Consequently, 
it means that considering the one-parameter family of diffeomorphisms which is generated by a vector field, one can define a quantity (tensor energy-momentum) which is
covariantly conserved $\p_\mu T^{\mu \nu} = 0$. The aforementioned result is an implementation of Noether's theorem to the case of Poincare group of symmetries.
It can be checked that on classical level the dilatation current defined as $j_{\mu ~\mathrm{(dil)}} = T_{\mu \nu} x^\nu$ is conserved, i.e., $\p^\mu j_{\mu ~\mathrm{(dil)}} =0$. 
It turns out that this current is not conserved at the quantum level~\cite{che22}.

\subsection{Quantum level}

The scale invariance is broken by quantum fluctuations \cite{che16} and on quantum level the dilatation current $ j_{\mu ~\mathrm{(dil)}} $ is not 
conserved, i.e., 
$\langle\p_\mu j^{\mu }_{\mathrm{(dil)}}\rangle =  {\langle T^{\alpha}_{\alpha}\rangle}\ne 0$ due to the quantum fluctuation which make the quantum average of $T^\alpha_\alpha$ non vanishing, as we show in what follows.
 
 In order to elaborate the problem in question,
 we assume that the {\it visible} and {\it dark matter} sectors have the same (dimensionful) renormalization scale $\mu$, and 
 restrict our considerations to the case when spacetime metric slightly changes from the Minkowski one, by a Weyl type conformal transformation 
 \be
 g_{\mu \nu} = e^{2 \Omega(x)} \eta_{\mu \nu},
\label{weyl}
 \ee  
 in which conformal factor is 
 much smaller than 1. It enables us to use method of quantum field theory in Minkowski spacetime and not to enter the area of quantum effects on curved spacetime background.

Technically, at the quantum level the scale invariance is not conserved, because of the fact that the coupling in gauge theories is a function
\cite{sch14} of renormalization scale $\mu$.


In our case we use $\beta$-functions for generalized gauge coupling $\te_A$ and $\te_B$, given by equations (\ref{cA})-(\ref{cB}), connected with two $U(1)$-gauge groups,
as well as, coupling $\alpha$ binding field strength tensors of {\it visible} and {\it hidden} sectors. 
 Denoting the changes of scale $\mu$ by $\delta \mu$ and taking the above metric transformation (\ref{weyl}) into account, we arrive at the following:
 \be
 \mu \rightarrow \mu + \delta \mu, \qquad \delta\mu = \mu ~\delta \Omega(x),
\ee
and the resulting modifications of  charges
\ben
 \te_A &\rightarrow &\te_A + \delta \te_A, \qquad  \te_A \rightarrow \te_B + \delta \te_B, \\ 
 \delta \te_A &=& \beta(\te_A) ~\delta\Omega, \qquad \delta \te_A = \beta(\te_A) ~\delta\Omega,
 \een
which we expressed {\it via} the scale factor $\Omega$.

 As we consider the transformation which slightly changes flat metric $\eta_{\mu \nu}$, i.e., for which $\mid \Omega (x) \mid \ll1$, one enables
  to rewrite the action for the system in question as follows:
  \be
  S_{\Omega} = S + \int d^4x~T_\beta{}^\beta~\Omega(x) + \cO(\Omega^2),
  \label{dil}
  \ee
  where we have denoted by  $S_{\Omega}$ the action for the transformed metric given by equation (\ref{weyl}) and by $S$ the action for Minkowski spacetime.
  Further, the above implies that the expectation value of the energy momentum tensor is given by the 
  functional~\cite{che16}
   \be
   \avr{T_\beta{}^\beta} =
   \frac{1}{i} \frac{1}{\cZ [\tA_\mu^{(cl)},~\tB_\mu^{(cl)}, \Omega] }~\frac{\delta 
  \cZ [\tA_\mu^{(cl)},~\tB_\mu^{(cl)}, \Omega] }{\delta \Omega},
   \ee
  for the generating function of our system defined by
  \be
{\cZ}[ \tA_\mu^{(cl)},~\tB_\mu^{(cl)},\Omega] = \int \cD \tA~ \cD \tB ~\cD \psi ~e^{i S_\Omega [\tA + \tA_{(cl)}, \tB + \tB_{cl}, {\bar \psi}, \psi]}.
\ee
One remarks that the system under consideration is coupled to the background classical gauge fields $\tA^{(cl)}_ \mu$ and $\tB^{(cl)}_\mu$.

As a result, the total average current is a sum of those related to {\it visible} and {\it dark} components. It is provided by
\ben
\avr{j^\mu (x)} &=& \avr{j^\mu(\tA_\beta)} + \avr{j^\mu(\tB_\ga)} \\ \nonumber
&=&  \frac{1}{i} \frac{1}{\cZ [\tA_\mu^{(cl)},~\tB_\mu^{(cl)}, \Omega] }~\frac{\delta 
  \cZ [\tA_\mu^{(cl)},~\tB_\mu^{(cl)}, \Omega] }{\delta \tA^{(cl)}_\mu} \\ \nonumber
  &+& \frac{1}{i} \frac{1}{\cZ [\tA_\mu^{(cl)},~\tB_\mu^{(cl)}, \Omega] }~\frac{\delta 
  \cZ [\tA_\mu^{(cl)},~\tB_\mu^{(cl)}, \Omega] }{\delta \tB^{(cl)}_\mu}.
 \een
Because of the fact that at quantum level gauge couplings and $\alpha$-coupling binding two sectors, are functions 
of the renormalization scale, the traces of the energy momentum tensors referred to gauge fields are not equal to zero.
 Moreover, taking into account the influence of the multi loop corrections,
as was shown in Ref. \cite{shi91}, lead to the following relations
\ben \label{eq:avrT}
\avr{T_\beta{}^{\beta} (\tA_\mu)} &=& \frac{\beta(\te_A)}{2 \te_A} \tF_{\mu \nu} \tF^{\mu \nu}, \qquad
\avr{T_\beta{}^{\beta} (\tB_\mu)} = \frac{\beta(\te_B)}{2 \te_B} \tB_{\mu \nu} \tB^{\mu \nu},\\
\avr{T_\beta {}^\beta }&=& \avr{T_\beta{}^{\beta} (\tA_\mu)} + \avr{T_\beta{}^{\beta} (\tB_\mu)} .
\een
In the above equations the field strength tensors are referred to classical background fields $\tA_\mu^{(cl)}$ and $\tB_\mu^{(cl)}$. For the brevity of 
the subsequent notation, one omits the superscript ${}^{(cl)}$ referring to the classical background field of {\it visible} and {\it hidden} sectors.

The current induced by small gauge fields pertaining both to {\it visible} and {\it dark sectors} and minute dilatations of the conformal factor $\Omega(x)$ is provided by 
\ben
\avr{j^\mu(x)} &=& \avr{j^\mu(x, \tA_\mu)}_{\mathrm{Kubo}} +  \avr{j^\mu(x, \tB_\mu)}_{\mathrm{Kubo}} 
+  \avr{j^\mu(x, \tA_\mu)}_{\mathrm{dilat}} +  \avr{j^\mu(x, \tB_\mu)}_{\mathrm{dilat}} \\ \nonumber
&+&
 \avr{j^\mu(x, \tA_\mu)}_{\mathrm{scale}} +  \avr{j^\mu(x, \tB_\mu)}_{\mathrm{scale}} + \dots
\een
where we have denoted the terms proportional to the first power of gauge fields $\tA_\mu$, $\tB_\mu$ and 
conformal factor $\Omega$, as well as, the to their products, i.e., $\tA_\mu \Omega,~\tB_\mu \Omega$, plus higher-order terms.

 The first two terms on the right-hand side of the above equation are given by ordinary Kubo relation calculated in a flat Minkowski spacetime, where we have
 the absence of external perturbations given by $\delta \tA_\mu =0,~\delta \tB_\mu = 0$ and $\delta g_{\mu \nu} =0$. The 
  ones proportional to scale factor
  are bounded with a linear response of the current
 to the pure dilation $S \rightarrow S_{\Omega}$, given by the relation (\ref{dil}). The same arguments as quoted in ordinary Maxwell case \cite{che16}, lead to the conclusion that 
it vanishes in the linear approximation regime so in the following section we deal with $scale$ term only.

\section{Scale current}
In our paper we pay attention to the terms connected with a scale anomalous contributions to the expectation value of the current influenced both by {\it visible Maxwell sector} and
{\it dark matter} one. Namely
\be
\avr{j^\mu}_{\mathrm{scale} } = \avr{j^\mu(x, \tA)}_{\mathrm{scale}} + \avr{j^\mu(x, \tB)} _{\mathrm{scale}}.
\ee
In order to find each contribution to $\avr{j^\mu}_{\mathrm{scale} } $ we use three-point functions connected with each of the considered sectors, i.e.,
\ben
\Pi^{\mu \nu} (x_i, \tA_\mu) &=& \avr{j^\mu(x, \tA_\mu)~j^\mu(y, \tA_\mu)~T_\beta{}^{\beta} (\tF_{\mu \nu},z)}  \mid_{{}_{\delta g_{\mu\nu} = 0}^{\tA_\mu = 0}}\\
&=& -\frac{\delta^2 \avr{T_\beta{}^{\beta} (\tF_{\mu \nu},z)}}{\delta \tA_\mu (x) \delta \tA_\nu(y)}  \mid_{{}_{g_{\mu\nu} \rightarrow \eta_{\mu\nu}}^{\tA_\mu \rightarrow 0}}, \\
\Pi^{\mu \nu} (x_i, \tB_\mu) &=& \avr{j^\mu(x, \tB_\mu)~j^\mu(y, \tB_\mu)~T_\beta{}^{\beta} (\tB_{\mu \nu},z)}  \mid_{{}_{\delta g_{\mu\nu} = 0}^{\tB_\mu = 0}}\\
&=& -\frac{\delta^2 \avr{T_\beta{}^{\beta} (\tB_{\mu \nu},z)}}{\delta \tB_\mu (x) \delta \tB_\nu(y)}  \mid_{{}_{g_{\mu\nu} \rightarrow \eta_{\mu\nu}}^{\tB_\mu \rightarrow 0}},
\een
and the direct calculations of functional derivatives of (\ref{eq:avrT})
reveal their explicit forms provided by
\be
\Pi^{ab} (x_i, \tA_\mu(\tB_\mu)) = - \frac{2 \beta(\te_A (\te_B))}{\te_A (\te_B)}~
\Big( \eta^{ab} ~\eta^{\mu \nu} - \eta^{\nu b}~ \eta^{\mu a} \Big)~\p_\mu \Big(\delta(z-y) \Big) \p_\nu \Big( \delta(z-x) \Big).
\label{kub}
\ee
Consequently, by virtue of the following 
 linear
response Kubo formulae:
\ben
 \avr{j^\mu(x, \tA)}_{\mathrm{scale}} &=& \int d^4y \int d^4z \Pi^{\mu \nu} (x_i, \tA_\mu)~\tA_\nu (y)~\Omega(z),\\
  \avr{j^\mu(x, \tB)}_{\mathrm{scale}} &=& \int d^4y \int d^4z \Pi^{\mu \nu} (x_i, \tB_\mu)~\tB_\nu (y)~\Omega(z),
\een
and equation (\ref{kub}), we arrive at the relations for anomalous currents generated by the Weyl scale anomaly
\ben \label{ja}
\avr{j^\mu(x, \tA)}_{\mathrm{scale}} &=&  \frac{2 \beta(\te_A)}{\te_A}\Big( -  \tF^{\mu \nu}~ \p_\nu \Omega (x) + \Omega(x) ~j^{\mu}_{\mathrm{class}}(\tA_\mu)\Big),\\ \label{jb}
\avr{j^\mu(x, \tB)}_{\mathrm{scale}} &=&  \frac{2 \beta(\te_B)}{\te_B}\Big( -\tB^{\mu \nu} ~\p_\nu \Omega (x) + \Omega(x)~ j^{\mu}_{\mathrm{class}}(\tB_\mu)\Big).
\een

Having in mind that in classical regime the equations of motion for the gauge sectors with adequate currents are of the forms
\be
\p_\mu \tF^{\mu \nu} = - j^{\mu}_{\mathrm{class}}(\tA_\mu), \qquad \p_\mu \tB^{\mu \nu} = - j^{\nu}_{\mathrm{class}}(\tB_\mu), 
\ee
one achieves the conservation of anomalously generated currents stemming from both sectors, i.e.,
\be
\p_\mu \avr{j^\mu}_{\mathrm{scale} } = \p_\mu \avr{j^\mu(x, \tA)}_{\mathrm{scale}} + \p_\mu\avr{j^\mu(x, \tB)} _{\mathrm{scale}} = 0.
\ee

In our considerations we are interested in the regime far from the classical one, therefore this fact will justify that we set classical currents of both sectors, in the 
region of dilatation equal to
zero. Relations (\ref{cA})-(\ref{cB}) and (\ref{ja})-(\ref{jb}), allow us to rewrite the equation for scale current in the form as follows:
\ben 
\avr{j^\mu}_{\mathrm{scale} } &=& \avr{j^\mu(x, \tA)}_{\mathrm{scale}} + \avr{j^\mu(x, \tB)} _{\mathrm{scale}} \\ \nonumber
 &=& - \frac{2}{e} F^{\mu \nu} \Big( \beta(\te_A) + \beta(\te_B) \Big) \p_v \Omega
 - \frac{2}{e} B^{\mu \nu} \Big( \beta(\te_A) - \beta(\te_B) \Big) \p_v \Omega.
 \een
 In the derivation of the above equation we have used the fact that
 \be
\tF_{\mu \nu} = \frac{\sqrt{2 -\alpha}}{2} \Big( F_{\mu \nu} - B_{\mu \nu}\Big),
\ee
and
\be
\tB_{\mu \nu} = \frac{\sqrt{2 +\alpha}}{2} \Big( F_{\mu \nu} + B_{\mu \nu}\Big),
\ee
which follows from the definitions (\ref{transA})-(\ref{transB}).
 
 On the other hand, the forms of time and spatial components of the strength gauge tensors
\ben \label{em}
E^{(F)}_i &=& F_{0i} = F^{i0}, \qquad B^{(F)}_k = - \frac{1}{2} \ep_{klm} F^{lm},\\ \label{emd}
E^{(B)}_i &=& F_{0i} = B^{i0}, \qquad B^{(B)}_k = - \frac{1}{2} \ep_{klm} B^{lm},
\een
authorize the following relations for the spatial components of the current in question:
\ben \label{jm}
\avr{j^m (x)}_{\mathrm{scale} } &=& - \frac{2}{e} \Big[ E^{(F) m} \Big( \beta(\te_A) + \beta(\te_B) \Big) 
+ E^{(B) m} \Big( \beta(\te_A) - \beta(\te_B) \Big) \Big] \p_0 \Omega(x) \\
 \nonumber
&+& \frac{2 }{e} ~\ep^{mjk} ~\p_j \Omega(x) ~\Big[ B^{(F)}_k \Big( \beta(\te_A) + \beta(\te_B) \Big) 
+ B^{(B)}_k  \Big( \beta(\te_A) - \beta(\te_B) \Big) \Big],
\een
while for the time component, one gets
\be
\avr{j^0(x)}_{\mathrm{scale} } = \frac{2}{e} E^{(F) m}  \Big( \beta(\te_A) + \beta(\te_B) \Big) \p_m \Omega(x)
+ \frac{2}{e} E^{(B) m}  \Big( \beta(\te_A) - \beta(\te_B) \Big) \p_m \Omega(x),
\label{jo}
\ee

Equations (\ref{jm})-(\ref{jo}) can be rewritten as follows:
\ben \label{c1}
\avr{j^m (x)}_{\mathrm{scale} } &=& \sigma_F (x_\mu)  E^{(F) m} + \sigma_B(x_\mu) E^{(B) m} \\ \nonumber
&+& \ep^{mlk} K_l^{(F)}  B^{(F)}_k + \ep^{mlk} K_l^{(B)}  B^{(B)}_k 
,\\ \label{c2}
\avr{j^0(x)}_{\mathrm{scale} } &=& K_m^{(F)}  E^{(F) m} + K_m^{(B)}~E^{(B) m}.
\een
 Note, that the scalar 
\be
\sigma_F (x_\mu) = - \frac{2}{e} ~\p_0 \Omega \Big( \beta(\te_A) + \beta(\te_B) \Big), \qquad
\sigma_B (x_\mu) = - \frac{2}{e} ~\p_0 \Omega \Big( \beta(\te_A) - \beta(\te_B) \Big),
\label{sss}
\ee
and vector parameters
\be
K_j^{(F)} = \frac{2}{e}~\p_j \Omega \Big( \beta(\te_A) + \beta(\te_B) \Big), \qquad
K_j^{(B)} = \frac{2}{e}~\p_j \Omega \Big( \beta(\te_A) - \beta(\te_B) \Big),
\label{kkk}
\ee
play the role of conductivities.

Having in mind equation (\ref{sss}), one can identify the quantity $\sigma_{A/B} (x_\mu)$ in the relation (\ref{c1}),
as a scale anomalous Ohm conductivity. For the case when the conformal scale factor is spatial uniform, i.e., $\p_j \Omega  =0$, one arrives at
\be
\avr{j^m (x)}_{\mathrm{scale} } = \sigma_F (x_\mu)  E^{(F) m} + \sigma_B(x_\mu) E^{(B) m},
\ee
and we call this phenomenon as a scale electric effect with {\it dark matter sector} (SEEDM).
It can be seen that {\it dark sector} influences Ohm-like  current, by $\sigma_B(x_\mu)$ and $ E^{(B) m}$.

The other situation takes place, when $\p_0 \Omega =0$. Then we obtain
\be
\avr{j^m (x)}_{\mathrm{scale} } = \ep^{mlk} K_l^{(F)}  B^{(F)}_k + \ep^{mlk} K_l^{(B)}  B^{(B)}_k 
\label{smedm}
\ee
where the spatial inhomogeneities of the conformal factor are included in the term given by (\ref{kkk}), when $\p_j \Omega \ne 0$.
One calls this phenomenon as 
 scale magnetic effect with {\it dark matter sector} (SMEDM).
Equation (\ref{smedm}) is connected with magnetic fields from both sectors and spatial inhomogeneities caused by $\p_j \Omega \ne 0$, being subject to the existence of two sector
beta-functions. As in the previous case (SEEDM), we get the {\it hidden sector} influence, proportional to {\it dark sector} magnetic field $ B^{(B)}_k$ and to $K_l^{(B)}$ 
bounded with {\it dark matter}
spatial  inhomogeneities.

On the other hand, inspection of the equation (\ref{c2}) reveals that because of the presence of electric fields (stemming from both sectors), the scale anomaly
induces the existence of $\avr{j^0 (x)}_{\mathrm scale}$ at spatial inhomogeneities of the spacetime.

\section{Possible applications}
In this section
 we discuss two possible sources of application of the above analysis. Firstly we calculate the modification of the metric by the passing gravitational wave. 
 Secondly one conducts the analysis of the effect of {\it dark photon} on the ohmic conductivities related to the visible and the {\it dark sectors} in the expanding Universe. 

\subsection{Gravitational wave passing through the system}
Gravitational waves can interact with matter in many ways due to the universality of gravitational coupling. For instance 
stressing and stretching it, being a mechanical coupling of gravitational waves. It give rise to the detection way like Weber bars, or interferometers where the wave couples to a resonant mass or mirrors. However exploring the searches of gravitational waves at higher frequencies (beyond 1 kHz) one ought to
consider the full set of gravitational couplings. There are two main approaches, i.e., exploring coupling of gravitational waves to electromagnetism or to the traditional
mechanical coupling adapted for short wavelength \cite{maggiore}, \cite{dom25}, \cite{wan23}.

In this section we shall consider linearized (weak) gravitational wave passing through the system under consideration and look for the possible effects
on derived expressions for the anomalous current.  The linearized gravity metric tensor  given by
\be
g_{\mu \nu} = \eta_{\mu \nu} + h_{\mu \nu}, \qquad \mid h_{\mu \nu} \mid \ll 1,
\ee
describes expansion of Einstein equations around the flat, Minkowski spacetime, to linear order in $h_{\mu \nu}$.

We will use freely falling frame (Riemann normal coordinates with the axes which are marked by gyroscopes), also named as Fermi normal coordinate system. 
 Therefore a freely falling frame is a local inertial frame along an entire studied geodesic. Practically it can be set out in a drag-free satellite, and the apparatus is
 in free fall in the total gravitational field coming from Earth and passing gravitational wave. If one considers a sufficiently small region of space we can choose
coordinates so that even in the presence of gravitational wave the metric is flat.
 
It turns out that  to the linear order in $\mid x_k \mid$,  one has no corrections to this metric, because of the fact that in a freely falling frame the derivatives of the 
metric tensor $g_{\mu \nu}$ disappear at the point of origin. Just
the expansion to second order, and the second derivatives of metric tensor written in term of the Riemann tensor gives the result, i.e.,
in the considered system, the line element describing the spacetime in the nearby of origin is given by the linear order in Riemann tensor \cite{ni78}
\be
ds^2 \simeq \Big( 1 + R_{i0j0} ~x^i x^j \Big) c^2 dt^2 - \Big( \delta_{ij} - \frac{1}{3} R_{imjr} ~x^m x^r \Big)dx^i dx^j + 2 \Big( \frac{2}{3} R_{0 m i r} ~x^m x^r \Big) cdt dx^i,
\label{grwav}
\ee
where the Riemann tensor is calculated at the origin of the system.

One ought to find the components of the Riemann tensor corresponding to a gravitational wave in the proper time detector frame. However, 
in TT-frame (transverse-traceless), where $h^{\mu 0}=0,~h_a{}^{a} =0,~\p^m h_{m j} =0$, it has the simplest form. As was shown e.g., in Ref. \cite{maggiore},
in the linearized theory of gravity Riemann tensor is invariant, i.e.,
for $h_{\mu \nu}$ we have that
\be
h_{\mu \nu}(x) \rightarrow h'_{\mu \nu}(x') = h_{\mu \nu} (x) - \Big( \p_\mu \xi_\nu + \p_\nu \xi_\mu \Big), \qquad
\mid \p_\mu \xi_\nu \mid \ll 1,~\mid h_{\mu \nu} \mid \ll 1,
\ee 
rather than covariant like in General Relativity. Just we can compute the Riemann tensor in TT-frame, because of its simplest form.
As a result the non-zero component of Riemann tensor yields
\be
R_{i0j0} = - \frac{1}{2 c^2} {\ddot h}^{TT}_{ij},
\label{curt}
\ee
with dots representing the derivative with respect to time.

For simplicity, let us consider
the metric of a weak gravitational wave going along $z$ axis,
which implies that the only non-zero components of the metric perturbations will be given by $h^{TT}_{xx} = - h^{TT}_{yy}$. As a result the only non-zero curvature tensor component in
the line element (\ref{grwav}) is described by (\ref{curt}). Thus the spacetime metric implies
\be
ds^2 \simeq \Big( 1  - \frac{1}{2 c^2} {\ddot h}^{TT}_{ij}~x^i x^j \Big) c^2 dt^2 - \delta_{ab} dx^a dx^b.
\ee
Recalling
that $\mid h_{ab} \mid \ll 1$ we can rewrite this metric in the form provided by the following line element:
\be
ds^2 \simeq \Big( 1  - \frac{1}{2 c^2} {\ddot h}^{TT}_{ij} ~x^i x^j \Big) \Big( c^2 dt^2 - \delta_{ij} dx^i dx^j \Big) + \cO( h^2),
\ee
thus the conformal factor implies
\be
\Omega(x) = \frac{1}{2} \ln \Big( 1  - \frac{1}{2 c^2} {\ddot h}^{TT}_{ij} ~x^i x^j \Big).
\ee
and
\be
\p_o \Omega  \simeq - \frac{1}{4c^2} {\dddot h}^{TT}_{ab}~x^a x^b + \cO(h^2), \qquad
\p_j \Omega \simeq - \frac{1}{2 c^2} {\ddot h}^{TT}_{aj}~x^a + \cO(h^2),
\label{grcoef}
\ee
Consequently, due to the convention of the previous chapters we should set $c=1$ in all these derivations.

The analysis addressed to the problem of weak gravitational wave passing through the considered system,
equations (\ref{grcoef}) and (\ref{sss})-(\ref{kkk}),
 reveals that it may be the ingredient of the
arousing factor of a scaling current. Both timelike and spacelike components stemming from {\it visible} and {\it hidden} sectors will be present.

\subsection{Expanding Universe}
 As a next example, we consider the case of expanding geometry, given by a homogeneous isotropic line element, being a model of Universe
\be
ds^2 = dt^2 - a(t)^2~\delta_{ij} dx^i dx^j,
\ee
where 
$a(t)$ is a scale factor connected with Universe expansion.
In order to apply the results obtained on the ground of quantum field theory in Minkowski background, the scale factor should be close to one.

Having in mind  he forms of $\beta$-functions (see appendix)
for the studied case
when we put $e_d =0$, the currents coefficients are provided by
\ben \nonumber
&{}&\avr{j^m (x)}_{\mathrm{scale} } = - \frac{\sqrt{2} e^2}{12 \pi^2 }
 \Big[ E^{(F) m} \Big( 1 + \frac{3}{8} \alpha^2 + \cO(\alpha^{2n}) \Big) + \frac{3}{4} \Big(\alpha +\cO(\alpha^{2n+1}))\Big)~E^{(B) m} \Big] \p_0 \Omega(x) \\
&+& \frac{\sqrt{2} e^2}{12 \pi^2 } \Big[ \ep^{mka} B_a^{(F)} \Big( 1 + \frac{3}{8} \alpha^2 + \cO(\alpha^{2n}) \Big)
+ \frac{3}{4} \Big(\alpha +\cO(\alpha^{2n+1}))\Big) \ep^{mka} B_a^{(F)} \Big] \p_k \Omega,
\een
for the spatial components and for timelike part one gets
\be
\avr{j^0 (x)}_{\mathrm{scale} } = - \frac{\sqrt{2} e^2}{12 \pi^2 }
 \Big[ E^{(F) k} \Big( 1 + \frac{3}{8} \alpha^2 + \cO(\alpha^{2n}) \Big) 
 + \frac{3}{4} \Big(\alpha +\cO(\alpha^{2n+1}))\Big)
 ~E^{(B) k} \Big] \p_k \Omega(x),
 \ee
where $n = 1,2,\dots$

The conductivity of massless Dirac field in the background of two coupled gauge fields charged under two $U(1)$-gauge groups, is given by two ingredients. One 
($\avr{j^m (x)}_{\mathrm{scale} }$)
stemming from the mixture of the 
{\it visible} and {\it dark sectors}, the other 
($\avr{j^0 (x)}_{\mathrm{scale} }$)
strictly bounded with {\it dark photon} sector. Up to the order of $\cO(\alpha^{n \ge3})$, they imply
\be
\sigma_F = - \frac{\sqrt{2} e^2}{12 \pi^2 } \Big( 1 + \frac{3}{8} \alpha^2 \Big) \frac{H(t)}{\hbar c},
\qquad
\sigma_{B} = - \frac{\sqrt{2} e^2}{16 \pi^2 }  \alpha~\frac{H(t)}{\hbar c},
\label{un}
\ee
where we have denoted by  $H(t) ={ \dot a}(t)/a(t)$
the Hubble parameter. One can observe that we have the influence of {\it dark sector} on conductivity bounded with $E_m^{(F)}$, which is quadratic in $\alpha$,
while
the {\it hidden sector} conductivity is proportional to $\alpha$.

Inspection of the relation (\ref{un}) reveals that the expressions reduces to the one found in Ref. \cite{che16}, when one excludes the influence of {\it dark matter}
and sets $\alpha = 0.$

\section{Conclusions}
In our paper we have elaborated the problem of the influence of gravitational background inhomogeneities on the emergence of anomalous currents.
We have considered the simple system of massless Dirac field in the background of Maxwell and {\it dark matter} fields. As the {\it dark sector} one takes into account
the massless
{\it dark photon} model, where the additional $U(1)$-gauge field is coupled to the {\it visible sector}.

The gravitational field inhomogeneities are modelled by Weyl type conformal transformation applied to the Minkowski line element. The conformal factor is supposed to be 
$\mid \Omega \mid  \ll 1$, which in turn enables us to restrict our considerations to the case of flat space quantum field theory, and not use apparatus of quantum field theory in curved background.
It simplifies calculations to a great extent.

We pay attention to the possible conductivity caused by passing weak gravitational wave through the considered system. Both conductivities stemming from {\it visible}
and {\it dark } sectors are effected by the gravitational wave and contribute to the scale current. In the case of time variation the time derivative of the scale factor is proportional to the third time derivative of $h_{ab}$ in transverse-traceless gauge,
while the gradient of conformal factor is connected with the second time derivative.

Moreover we have found that in expanding geometry described by the line element of inflating flat Universe, the conductivity of massless charged fermions influenced by the 
{\it dark sector} has additional term proportional to $\alpha^2$. Additionally the conductivity bounded with {\it dark photon} is proportional to
the kinetic mixing parameter $\alpha$ and Hubble's parameter. It remains to be seen if this contribution to the scale current could be detected in future clever experiments.

One of the possible ways of finding experimental evidence of the {\it dark photon} imprinted 
in the scale anomaly currents, could be to look for thermoelectric effects of this phenomenon in Weyl and Dirac
semimetals in a close analogy to the proposal  \cite{che18}-\cite{che22}. In these papers, it has been shown that the conformal anomaly
generates a bulk electric current perpendicular to a temperature gradient and the direction of
magnetic field background.

\section{Appendix}

\subsection{Calculation of the beta-functions}

Bare Lagrangian for the studied theory of {\it dark photon} yields
\be
\cL = - \frac{1}{4} \tF_{\mu \nu}^{(0)}\tF^{\mu \nu (0)} - \frac{1}{4} \tB_{\mu \nu}^{(0)}\tB^{\mu \nu (0)} 
+ {\bar \psi}^{(0)} \Big( i ~\psl -\te_A ^{(0)}~\tAsl^{(0)} -\te_B^{(0)}~\tBsl^{(0)} \Big) \psi^{(0)},
\label{bare}
\ee
where the dimensions of the fields are as follows:
\be
[\tA_\mu^{(0)}] = [\tB_\mu^{(0)}] = \frac{d-2}{2}, \quad
[\psi^{(0)}] = \frac{d-1}{2}, \quad
[\te_A] = [\te_B] = \frac{4-d}{2}.
\ee
The quantities in (\ref{bare}) are finite, or for $(d=4-\ep)$-dimensional case they are finite but scale as inverse powers of $\ep$. The bare charges are dimensionless
if $d=4$.

Our next task is to express the above Lagrangian in terms of physical renormalized fields and charges.
As has mentioned before, in the regime under consideration, we shall neglect the running $\alpha$.

 We want to obtain
the charges $\te_A$ and $\te_B$ as numbers, fields should have canonical normalization.
Consequently, the aforementioned requirements imply that the 
 fields and charges
in (\ref{bare}) rescale as 
\ben
\tA_\mu &=& \frac{1}{\sqrt{Z_3}} \tA_\mu^{(0)}, \qquad
\tB_\mu = \frac{1}{\sqrt{\tZ_3} }\tB_\mu^{(0)}, \\
\psi &=& \frac{1}{\sqrt{Z_2}} \psi^{(0)},\\
\te_A &=& \frac{1}{Z_{\te_A}} \mu^{\frac{d-4}{2}}~\te_A^{(0)}, \qquad \te_B = \frac{1}{Z_{\te_B}} \mu^{\frac{d-4}{2}}~\te_B^{(0)},
\een
where $\te_A,~\te_B$ and $Z_j$ are dimensionless.

The aforementioned procedure leads to the Lagrangian of the form
\ben
\cL &=& - \frac{1}{4} Z_3~\tF_{\mu \nu} \tF^{\mu \nu} - \frac{1}{4} \tZ_3~\tB_{\mu \nu} \tB^{\mu \nu} + Z_2{ \bar \psi} ~i ~\psl \psi \\ \nonumber
&-& Z_{1 \tA}~{\bar \psi}~ \tAsl \psi ~\te_A - Z_{1 \tB}~{\bar \psi}~ \tBsl \psi ~\te_B,
\een
where we set
\be
Z_{1 \tA} = Z_2~ \sqrt{Z_3} ~Z_{\te_A}, \qquad
Z_{1 \tB} = Z_2~ \sqrt{\tZ_3} ~Z_{\te_B}.
\ee
Having in mind one-loop corrections counter-terms $\delta_j$ \cite{sch14}, one has that for each of them, we have the following definitions:
\be
Z_a = 1 + \delta_a.
\ee
$Z_{1 \tA}$ and $Z_{1 \tB}$ stem from the vertex of the adequate gauge field and fermions, $Z_2$ is connected with the fermion self-energy, while
$Z_3$ and $\tZ_3$ are bounded with vacuum polarization diagrams for respectively $\tA_\mu$ and $\tB_\mu$.
 As was shown in \cite{sch14}, for $U(1)$-gauge theory $Z_1 (gauge~field) =Z_2$,  then in our case $\beta$-function can be calculated for 
$Z_3$ and $\tZ_3$ functions.
The form of the diagrams will be the same as for the ordinary QED, with $\tA_\mu,~\tB_\mu,~\te_A,~\te_B$ (see, e.g., \cite{sch14}).

Because of the fact that there is no $\mu$-dependence in bare Lagrangian, but in 
the renormalized one, accordingly we get
\be
\mu ~\frac{d}{d \mu} \te_A^{(0)} = 0, \qquad \mu ~\frac{d}{d \mu} \te_B^{(0)} = 0.
\ee
Inspection of the above relations at leading order in the expansion, namely for $Z_{1 \te_A, \te_B} =1$ and next leading order reveals the forms
of $\beta$-functions for charges $\te_A$ and $\te_B$, as follows:
\ben
\beta (\te_A) &=& -\frac{\ep}{2} \te_A + \frac{\te_A^3}{12 \pi^2} + \dots, \\
\beta (\te_B) &=& -\frac{\ep}{2} \te_B + \frac{\te_B^3}{12 \pi^2} + \dots,
\een
where we have denoted $\ep = 4 -d$.  We should take the limit $\ep \rightarrow 0$ in dimensional regularization.\\

\acknowledgements
M.R and K.I.W were partially supported by Grant No. 2022/45/B/ST2/00013 of the National Science Center, Poland.  



\begin{thebibliography}{99}

%
\def\cmp#1#2#3#4{\emph{#4}, \emph{ Commun. Math. Phys.} {\bf #1} (#3) #2}
\def\lmp#1#2#3#4{\emph{#4}, \emph{ Lett. Math. Phys.} {\bf #1} (#3) #2}
\def\hpa#1#2#3#4{\emph{#4}, \emph{ Hell. Phys. Acta} {\bf #1} (#3) #2}
\def\grg#1#2#3#4{\emph{#4}, \emph{ Gen. Rel. Grav.} {\bf #1} (#3) #2}
\def\pr#1#2#3#4{\emph{#4}, \emph{ Phys. Rev.} {\bf #1} (#3) #2}
\def\prl#1#2#3#4{\emph{#4}, \emph{ Phys. Rev. Lett.} {\bf #1}, #2 (#3)}
\def\prd#1#2#3#4{\emph{#4}, \emph{ Phys. Rev. D} {\bf #1}, #2 (#3)}

\def\prb#1#2#3#4{\emph{#4}, \emph{ Phys. Rev. B} {\bf #1}, #2 (#3) }
\def\prx#1#2#3#4{\emph{#4}, \emph{ Phys. Rev. X} {\bf #1} (#3) #2}
\def\pl#1#2#3#4{\emph{#4}, \emph{ Phys. Lett.} {\bf #1} (#3) #2}
\def\pla#1#2#3#4{\emph{#4}, \emph{ Phys. Lett. A} {\bf #1} (#3) #2 }
\def\plb#1#2#3#4{\emph{#4}, \emph{ Phys. Lett. B} {\bf #1}, #2 (#3)}
\def\prep#1#2#3#4{\emph{#4}, \emph{ Phys. Reports} {\bf #1}, #2 (#3)}
\def\phys#1#2#3#4{\emph{#4}, \emph{ Physica} {\bf #1} (#3) #2}
\def\jcp#1#2#3#4{\emph{#4}, \emph{ J. Comput. Phys.} {\bf #1} (#3) #2}
\def\jmp#1#2#3#4{\emph{#4}, \emph{ J. Math. Phys.} {\bf #1} (#3) #2}
\def\jpm#1#2#3#4{\emph{#4}, \emph{ J. Phys. A: Math. Gen.} {\bf #1} (#3) #2}
\def\cpr#1#2#3#4{\emph{#4}, \emph{ Computer Phys. Rept.} {\bf #1} (#3) #2}
\def\cqg#1#2#3#4{\emph{#4}, \emph{ Class. Quant. Grav.} {\bf #1} (#3) #2}
\def\cma#1#2#3#4{\emph{#4}, \emph{ Computers Math. Applic.} {\bf #1} (#3) #2}
\def\mc#1#2#3#4{\emph{#4}, \emph{ Math. Compt.} {\bf #1} (#3) #2}
\def\apj#1#2#3#4{\emph{#4}, \emph{ Astrophys. J.} {\bf #1} (#3) #2}
\def\apjs#1#2#3#4{\emph{#4}, \emph{ Astrophys. J. Suppl.} {\bf #1} (#3) #2}
\def\apjl#1#2#3#4{\emph{#4}, \emph{ Astrophys. J. Lett.} {\bf #1} (#3) #2}
\def\acta#1#2#3#4{\emph{#4}, \emph{ Acta Astronomica} {\bf #1} (#3) #2}
\def\apl#1#2#3#4{\emph{#4}, \emph{ Ann. Physik. (Leipzig)} {\bf #1} (#3) #2}
\def\amjp#1#2#3#4{\emph{#4}, \emph{Am. J. Phys.} {\bf #1} (#3) #2}
\def\anp#1#2#3#4{\emph{#4}, \emph{ Ann. Phys.} {\bf #1} (#3) #2}
\def\sa#1#2#3#4{\emph{#4}, \emph{ Sov. Astro.} {\bf #1} (#3) #2}
\def\sia#1#2#3#4{\emph{#4}, \emph{ SIAM J. Sci. Statist. Comput.} {\bf #1} (#3) #2}
\def\aa#1#2#3#4{\emph{#4}, \emph{ Astron. Astrophys.} {\bf #1} (#3) #2}
\def\mnras#1#2#3#4{\emph{#4}, \emph{ Mon. Not. R. Astr. Soc.} {\bf #1} (#3) #2}
\def\npb#1#2#3#4{\emph{#4}, \emph{ Nucl. Phys. B} {\bf #1}, #2 (#3)}
\def\npa#1#2#3#4{\emph{#4}, \emph{ Nucl. Phys. A} {\bf #1} (#3) #2}

\def\prsla#1#2#3#4{\emph{#4}, \emph{ Proc. R. Soc. London, Ser. A} {\bf #1} (#3) #2}
\def\jhep#1#2#3#4{\emph{#4}, \emph{ JHEP} {\bf #1} (#2) #3}
\def\jcap#1#2#3#4{\emph{#4}, \emph{ JCAP} {\bf #1} (#2) #3}

\def\nuca#1#2#3#4{\emph{#4}, \emph{ Nuovo Cimento A } {\bf #1} (#3) #2}
\def\nucb#1#2#3#4{\emph{#4}, \emph{ Nuovo Cimento B } {\bf #1} (#3) #2}
\def\ijmp#1#2#3#4{\emph{#4}, \emph{ Int. J. Mod. Phys. D} {\bf #1} (#3) #2}
\def\atmp#1#2#3#4{\emph{#4}, \emph{ Adv. Theor. Math. Phys.} {\bf #1} (#3) #2}
\def\ptps#1#2#3#4{\emph{#4}, \emph{ Prog. Theor. Phys. Suppl.} {\bf #1} (#3) #2}
\def\ptp#1#2#3#4{\emph{#4}, \emph{ Prog. Theor. Phys.} {\bf #1} (#3) #2}
\def\lmp#1#2#3#4{\emph{#4}, \emph{ Lett. Math. Phys.} {\bf #1} (#3) #2}
\def\cpam#1#2#3#4{\emph{#4}, \emph{ Comm. Pure Appl. Math.}  {\bf #1} (#3) #2}
\def\adv#1#2#3#4{\emph{#4}, \emph{ Adv. Phys.}  {\bf #1} (#3) #2}
\def\zh#1#2#3#4{\emph{#4}, \emph{ Zh. Eksp. Teor. Fiz.}  {\bf #1} (#3) #2}
\def\mplb#1#2#3#4{\emph{#4}, \emph{ Mod. Phys. Lett. B} {\bf #1} (#3) #2}
\def\jams#1#2#3#4{\emph{#4}, \emph{ J. Austral. Math. Soc. B} {\bf #1} (#3) #2}
\def\appa#1#2#3#4{\emph{#4}, \emph{ Acta Phys. Polonica A} {\bf #1} (#3) #2}
\def\appb#1#2#3#4{\emph{#4}, \emph{ Acta Phys. Polonica B} {\bf #1} (#3) #2}

\def\nat#1#2#3#4{\emph{#4}, \emph{Nature} {\bf #1} #2 (#3)}
\def\natcom#1#2#3#4{\emph{#4}, \emph{Nature Commun.} {\bf #1} (#3) #2}
\def\natphys#1#2#3#4{\emph{#4}, \emph{Nature Physics} {\bf #1} (#3) #2}
\def\natmat#1#2#3#4{\emph{#4}, \emph{Nature Mat.} {\bf #1} (#3) #2}


\def\science#1#2#3#4{\emph{#4}, \emph{Science} {\bf #1} (#3) #2}
\def\sciadv#1#2#3#4{\emph{#4}, \emph{Sci. Adv.} {\bf #1} (#3) #2}

\def\arcmp#1#2#3#4{\emph{#4}, \emph{Annual Rev. of Cond. Matter Physics} {\bf #1} (#3) #2}
\def\zphys#1#2#3#4{\emph{#4}, \emph{Z. Phys.} {\bf #1}, (#3) #2}
\def\ncs#1#2#3#4{\emph{#4}, \emph{Nuovo Cimento Suppl.} {\bf #1} (#3) #2}
\def\physb#1#2#3#4{\emph{#4}, \emph{Physica B} {\bf #1}, (#3) #2}
\def\jpcm#1#2#3#4{\emph{#4}, \emph{J. Phys.: Condens. Matter } {\bf #1} (#3) #2}
\def\pnas#1#2#3#4{\emph{#4}, \emph{Proc. Nat. Academy Sciences} {\bf #1} (#3) #2}
\def\sssr#1#2#3#4{\emph{#4}, \emph{Izv. Akad Nauk SSSR, ser. fiz.} {\bf #1} (#3) #2}
\def\jpg#1#2#3#4{\emph{#4}, \emph{ J. Phys. G} {\bf #1} (#3) #2}
\def\chinpb#1#2#3#4{\emph{#4}, \emph{Chin. Phys. B} {\bf #1} (#3) #2}
\def\njp#1#2#3#4{\emph{#4}, \emph{ New J. Phys.} {\bf #1} (#3) #2}
\def\frontphys#1#2#3#4{\emph{#4}, \emph{ Front. Phys.} {\bf #1} (#3) #2}
\def\epl#1#2#3#4{\emph{#4}, \emph{ EPL} {\bf #1} (#3) #2}
\def\rmp#1#2#3#4{\emph{#4}, \emph{ Rev. Mod. Phys.} {\bf #1}, #2 (#3)}
\def\rpp#1#2#3#4{\emph{#4}, \emph{ Rep. Prog. Phys.} {\bf #1}, #2 (#3)}

\def\hepph#1#2{{ hep-ph }{#1} (#2)}
\def\arxiv#1#2#3{\emph{#3},{ arXiv }{#1} (#2)}
\def\hepth#1#2{{ hep-th }{#1} (#2)}
\def\grqc#1#2{{ gr-qc }{#1} (#2)}
\def\ibid#1#2#3#4{\emph{#4}, {\it ibid.} {\bf #1} (#3) #2}
\def\conphy#1#2#3#4{\emph{#4}, \emph{Contemporary Physics} {\bf #1}, (#3) #2}
\def\ppnp#1#2#3#4{\emph{#4}, \emph{ Prog. Part. Nucl. Phys} {\bf #1} (#3) #2}
\def\arnps#1#2#3#4{\emph{#4}, \emph{ Annu. Rev. Nucl. Part. Sci.} {\bf #1} (#3) #2}
\def\ijmpa#1#2#3#4{\emph{#4}, \emph{ Int. J. Mod. Phys. A} {\bf #1}, #2 (#3)}
\def\jams#1#2#3#4{\emph{#4}, \emph{ J. Austral. Math. Soc. B} {\bf #1} (#3) #2}
\def\appa#1#2#3#4{\emph{#4}, \emph{ Acta Phys. Polonica A} {\bf #1}, (#3) #2}
\def\nat#1#2#3#4{\emph{#4}, \emph{Nature} {\bf #1}, (#3) #2}
\def\science#1#2#3#4{\emph{#4}, \emph{Science} {\bf #1}, (#3) #2}
\def\arcmp#1#2#3#4{\emph{#4}, \emph{Annual Rev. of Cond. Matter Physics} {\bf #1}, (#3) #2}
\def\jcap#1#2#3#4{\emph{#4}, \emph{JCAP} {\bf #1}, (#3) #2}
\def\conphy#1#2#3#4{\emph{#4}, \emph{Contemporary Physics} {\bf #1}, (#3) #2}
\def\ptps#1#2#3#4{\emph{#4}, \emph{ Prog. Theor. Phys. Suppl.} {\bf #1} (#3) #2}
\def\ptp#1#2#3#4{\emph{#4}, \emph{ Prog. Theor. Phys.} {\bf #1} (#3) #2}
\def\apjsup#1#2#3#4{\emph{#4}, \emph{ Astrophys. J. Suppl. Ser.} {\bf #1} (#3) #2}
\def\eurphysjc#1#2#3#4{\emph{#4}, \emph{ Eur. Phys. J.  C} {\bf #1}, #2 (#3)}
\def\njp#1#2#3#4{\emph{#4}, \emph{ New J. Phys. } {\bf #1} (#3) #2}
\def\eurphysjplus#1#2#3#4{\emph{#4}, \emph{ Eur. Phys. J.  Plus} {\bf #1}, #2 (#3)}
%
\def\hepph#1#2{{ hep-ph }{#1} (#2)}
\def\hepth#1#2{{ hep-th }{#1} (#2)}
\def\astroph#1#2{{ astro-ph }{#1} (#2)}
\def\grqc#1#2{{ gr-qc }{#1} (#2)}
\def\ibid#1#2#3#4{\emph{#4}, {\it ibid.} {\bf #1} (#3) #2}


\def\contp#1#2#3#4{\emph{#4}, \emph{ Contemporary Physics} {\bf #1}, #2 (#3)}
\def\revphys#1#2#3#4{\emph{#4}, \emph{Reviews in Phys.} {\bf #1} #2 (#3) }
\def\cag#1#2#3#4{\emph{#4}, \emph{ Commun. Anal. Geom.} {\bf #1} #2 (#3) }
\def\contmath#1#2#3#4{\emph{#4}, \emph{ Contemp. Math.} {\bf #1} #2 (#3)}
\def\epjc#1#2#3#4{\emph{#4}, \emph{ Eur. Phys. J. C} {\bf #1} #2 (#3) }
\def\revphys#1#2#3#4{\emph{#4}, \emph{Reviews in Phys.} {\bf #1} #2 (#3) }
\def\rmp#1#2#3#4{\emph{#4}, \emph{ Rev. Mod. Phys.} {\bf #1}, #2 (#3)}


%







\bibitem{neu11}
D. E. Neuenschwander, {\it Emmy Norther's wonderful theorem}, The John Hopkins University Press, Baltimore (2011).

\bibitem{ber96}
R. A. Bertlmann, {\it Anomalies in Quantum Field Theory}, Oxford, UK,
Clarendon, 1996, International Series of Monographs on Physics: \textbf{91}.
, 1996, p. 566. 


\bibitem{shi91}
M. A. Shifman, \prep{209}{341}{1991}{Anomalies in gauge theories}.

\bibitem{lan16}
K. Landsteiner, {\it Notes on anomaly induced transport}, Acta Phys. Polon. 47 (2016) 2617.


\bibitem{son04}
D. T. Son and A. R. Zhitnisky, \prd{70}{07018}{2004}{Quantum anomalis in dense matter}.
\bibitem{met05}
M. A. Metlitski and A. R. Zhitnisky, \prd{72}{045011}{2005}{Anomalous axion interactions and topological currents in dense matter}.
\bibitem{fuk08}
K. Fukushima, D. E. Kharzeev, and H. J. Warringa, \prd{78}{074033}{2008}{The chiral magnetic effect}.
\bibitem{vil80}
A. Vilenkin, \prd{22}{3080}{1980}{Macroscopic party violation current in a magnetic field}.
\bibitem{vil79}
A. Vilenkin, \prd{20}{1807}{1979}{Macroscopic parity violating effects: Neutrino fluxes from rotating black holes and in rotating thermal radiation}.
\bibitem{son09}
D. T. Son and P. Sur\'owka, \prl{103}{191601}{2009}{Hydrodynamics with triangle anomalies}.
\bibitem{lan11}
K. Landsteiner, E. Megias, and F. Pena-Benitez, \prl{107}{021601}{20111}{Gravitational anomaly and transport}.

\bibitem{hyd}
R. Baier, P. Romatschke, D. T. Son, A. O. Starinets, and M. A. Stephanov, \jhep{04}{2008}{100}{Relativistic viscous hydrodynamics, conformal invariance, and holography},
~R. Loganayagam, \jhep{05}{2008}{087}{Entropy current in conformal hydrodynamics}.





\bibitem{che16}
M. N. Chernodub, \prl{117}{141601}{2016}{Anomalous transport due to the conformal anomaly}.
\bibitem{che18}
M. N. Chernodub, A. Cortijo, and M. A. H. Vozmediano, \prl{120}{206601}{2018}{Generation of a Nerst current from the conformal anomaly in Dirac and Weyl semimetals}.
\bibitem{che22}
M. N. Chernodub, Y. Ferreiros, A. G. Grushin, K. Landsteiner,
M. A.H. Vozmediano, \prep{977}{158}{2022}{Thermal transport, geometry, and anomalies}.


\bibitem{hol86}
B. Holdom, \plb{166}{196}{1986}{Two $U(1)$'s and $\ep$ charge shifts}.
\bibitem{cap21}
A. Caputo, A. J. Millar, C. A. J. O'Hare, and E. Vitagliano, \prd{104}{095029}{2021}{Dark photon limits: A handbook}.

\bibitem{ach16}
B.S. Acharya, S.A.R. Ellis, G.L. Kane, B.D. Nelson, and M.J. Perry, \prl{117}{181802}{2016}{Lightest Visible-Sector Supersymmetric Particle is Likely Unstable}.

\bibitem{gra16}
P.W. Graham, J. Mardon, and S. Rajendran, \prd{93}{103520}{2016}{Vector dark matter from inflationary fluctuations}
\bibitem{sat22}
T. Sato, F. Takahashi, and M. Yamada, \jcap{08}{2022}{022}{Gravitational production of dark photon dark matter with mass generated by the Higgs mechanism}

\bibitem{axionosc}
P. Agrawal, N. Kitajima, M. Reece, T. Sekiguchi, and F. Takahashi, \plb{801}{135136}{2020}{Relic abundance of dark photon dark matter},\\
 R.T. Co, A. Pierce, Z. Zhang, and Y. Zhao, \prd{99}{075002}{2019}{Dark photon dark matter produced by axion oscillations}, \\
M. Bastero-Gil, J. Santiago, L. Ubaldi, and R. Vega- Morales, \jcap{04}{2019}{015}{Vector dark matter production at the end of inflation}.
\bibitem{dro19}
J.A. Dror, K. Harigaya, and V. Narayan, \prd{99}{035036}{2019}{Parametric resonance production of ultralight vector dark matter}.

\bibitem{reheat}
A. Ahmed, B. Grzadkowski, and A. Socha, \jhep{08}{2020}{059}{Gravitational production of vector dark matter},\\
Y. Ema, K. Nakayama, and Y. Tang, \jhep{09}{2018}{135}{
Production of purely gravitational dark matter}.

\bibitem{cstrings}
 A.J. Long and L.-T. Wang, \prd{99}{063529}{2019}{ Dark photon dark matter from a network of cosmic strings},\\
 N. Kitajima and K. Nakayama, {\it Dark photon dark matter from cosmic strings and gravitational wave background}, \hepth{2212.13573}{2022}.






\bibitem{jea03}
P. Jean {\it et al.}, \aa{407}{L55}{2003}{Early SPI/INTEGRAL measurements of 511 keV line emission from the 4th quadrant of the Galaxy}.
\bibitem{cha08}
J. Chang {\it et al.}, \nat{456}{362}{2008}{An excess of cosmic ray electrons at energies of 300-800 GeV}.
\bibitem{bub14}
E. Bulbul et al., \apj{789}{13}{2014}{Detection of an unidentified emission line in the stacked X-ray spectrum of galaxy clusters}.

\bibitem{fil20}
A. Filippi and M. De Napoli, \revphys{5}{100042}{2020}{Searching in the dark: the hunt for the dark photon}.
\bibitem{ger15}
A. Geringer-Sameth and M.G. Walker, \prl{115}{081101}{2015}{Indication of Gamma-Ray Emission from the Newly Discovered Dwarf Galaxy Reticulum II}.
\bibitem{bod15}
K.K. Boddy and J. Kumar, \prd{92}{023533}{2015}{Indirect detection of {\it dark matter} using MeV-range gamma-rays telescopes}.
\bibitem{til15}
K.Van Tilburg, N. Leefer, L. Bougas, and D. Budker, \prl{115}{011802}{2015}{Search for Ultralight Scalar Dark Matter with Atomic Spectroscopy}.

\bibitem{cha17}
J.H. Chang, R. Essig, and S.D. McDermott, \jhep{01}{2017}{107}{Revisiting Supernova 1987A constraints on dark photons}.
\bibitem{sensei}
M. Crisler et. al. (SENSEI Collaboration), \prl{121}{061803}{2019}{SENSEI: First Direct-Detection Constraints on Sub-GeV Dark Matter from a Surface Run}
\bibitem{lee14}
J.P. Lees et al., \prl{113}{201801}{2014}{Search for a Dark Photon in $e^+ e^-$ Collisions at BABAR}.
\bibitem{dav11}
M. Davier et al., \epjc{71}{1515}{2011}{Reevaluation of the hadronic contributions to the muon g-2 and to $\alpha(M^2_z)$}. 



\bibitem{fil23}
M. Filzinger, S. D\"orscher, R. Lange, J. Klose, M. Steinel, E. Benkler, E. Peik, C. Lisdat, and N. Huntemann, \prl{130}{253001}{2023}
{Improved Limits on the Coupling of Ultralight Bosonic Dark Matter to Photons from Optical Atomic Clock Comparisons}.
\bibitem{ram23}
K. Ramanathan, N. Klimovich, R. Basu Thakur, B.H. Eom, H.G. Leduc, S. Shu, A.D. Beyer, and P.K. Day,
\prl{130}{231001}{2023}{Wideband Direct Detection Constraints on Hidden Photon Dark Matter with the QUALIPHIDE Experiment}.
\bibitem{kot23}
 S. Kotaka, S. Adachi, R. Fujinaka, S. Honda, H. Nakata, Y. Seino, Y. Sueno, T. Sumida, J. Suzuki, O. Tajima, and S. Takeichi, \prl{130}{071805}
 {2023}{Search for Dark Photon Dark Matter in the Mass Range $74-110\text{ }\text{ }\mathrm{\ensuremath{\mu}}\mathrm{eV}$ with a Cryogenic Millimeter-Wave Receiver}.


\bibitem{fab21}
M. Fabbrichesi, E. Gabrielli, and G. Lanfranchi, {\it The Physics of the Dark Photon - a Primer}, Springer 2021.
\bibitem{an20}
H. An, M. Pospelov, J. Pradler, and A. Ritz, \prd{102}{115022}{2020}{New limits on dark photons from solar emission and keV scale dark matter}.
\bibitem{an13}
H. An, M. Pospelov, and J. Pradler, \plb{725}{190}{2013}{New stellar constraints on dark photon}.



\bibitem{sch14}
M. E. Peskin, D. V. Schroeder, {\it An Introduction to Quantum Field Theory}, Addison-Wesley, Reading, USA, 1995,~
M. D. Schwartz, {\it Quantum Field Theory and the Standard Model}, Cambridge University Press, Cambridge 2014,~
A. G. Williams, {\it Introduction to Quantum Field Theory -Classical Mechanics to Gauge Field Theories}, Cambridge University Press, Cambridge 2023.






\bibitem{maggiore}
M. Maggiore, {\it Gravitational Waves, vol. I, Theory and Experiments}, Oxford University Press, Oxford 2008.
\bibitem{dom25}
V. Domcke, S. A. R. Ellis, and N. L. Rodd, \prl{134}{231401}{2025}{Magnets are Weber bar gravitational wave detectors}.
\bibitem{wan23}
N. Wanwieng, N. Chattrapiban, A. Watcharangkool, \cqg{40}{235004}{2023}{The effects of gravitational waves on a hydrogen atom}.


\bibitem{ni78}
W-T. Ni and M. Zimmermann, \prd{17}{1473}{1978}{Inertial and gravitational effects in the proper reference fame of an accelerated, rotating observer}.






\end{thebibliography}
\end{document}